\RequirePackage{lineno}
\documentclass[12pt,onecolumn]{article}	
\usepackage[top=1in, bottom=1in, left=1in, right=1in]{geometry}
\usepackage{graphicx}
\usepackage{amssymb}
\usepackage{amsmath}  				
\usepackage{graphicx,float,wrapfig} 	
\usepackage{color}					
\usepackage[colorlinks,linktocpage=true]{hyperref}		
\usepackage[font=small,labelfont=bf,textfont=sf,
labelsep=quad]{caption}				
\usepackage{titlesec}

\titleformat{\section}{\normalfont\fontsize{14}{15}\bfseries}{\thesection}{1em}{}
\titleformat{\subsection}{\normalfont\fontsize{12}{15}\bfseries}{\thesubsection}{1em}{}
\titleformat{\subsubsection}{\normalfont\fontsize{11}{15}\bfseries}{\thesubsubsection}{1em}{}

\usepackage{gensymb}
\usepackage{caption}
\usepackage{subfig}

\hypersetup{						
  citecolor= black,
  linkcolor=blue,
}

\usepackage{setspace}   

\usepackage [english]{babel}
\usepackage [english = american]{csquotes}
\MakeOuterQuote{"}

 \usepackage[protrusion=false,expansion=false]{microtype}
  
 \usepackage[colorinlistoftodos]{todonotes}
\usepackage{soul}

\let\baraccent=\= 
\renewcommand{\=}[1]{\stackrel{#1}{=}} 

\begin{document}
\begin{center}
\Large{\textbf{Generation of droplet arrays with rational number spacing patterns driven by a periodic energy landscape}}\\
\end{center}
\noindent
\begin{center}
Anatoly Rinberg$^1$, Georgios Katsikis$^2$\& Manu Prakash$^{3\ast}$\\
\small{$^{1}$School of Engineering and Applied Sciences, Harvard University, Cambridge, Massachusetts 02138, USA, $^{2}$Department of Mechanical Engineering, Stanford University, Stanford, California 94305, USA} \\
\small{$^{3}$Department of Bioengineering, Stanford University, Stanford, California 94305, USA} \\
\small{$^\ast$To whom correspondence should be addressed; E-mail:  trinberg@gmail.com}
\end{center}




\begin{spacing}{1.5}
\noindent\textbf{Abstract}\\
\footnotesize
The generation of droplets at low Reynolds numbers is driven by non-linear dynamics that give rise to complex patterns concerning both the droplet-to-droplet spacing and the individual droplet sizes. Here we demonstrate an experimental system in which a time-varying energy landscape provides a periodic magnetic force that generates an array of droplets from an immiscible mixture of ferrofluid and silicone oil. The resulting droplet patterns are periodic, owing to the nature of the magnetic force, yet the droplet spacing and size can vary greatly by tuning a single bias pressure applied on the ferrofluid phase; for a given cycle period of the magnetic force, droplets can be generated either at integer multiples (1, 2, etc.), or at rational fractions (3/2, 5/3, 5/2, etc.) of this period with mono- or multidisperse droplet sizes. We develop a discrete-time dynamical systems model not only to reproduce the phenotypes of the observed patterns but also provide a framework for understanding systems driven by such periodic energy landscapes.
\end{spacing}
\doublespacing
\normalsize
\newpage
\section*{Introduction}
\label{sec:intro}
Discrete-time dynamical systems have been used to study physical phenomena such as population dynamics of predator-prey behavior \cite{neubert1995dispersal, liu2007complex}, spatial ecological patterns \cite{kot1992discrete}, control theory \cite{ogata1995discrete} and chaotic electronic circuits \cite{campos2009simple}. These systems often deal with recursive mathematical relations and use iterative maps to describe behaviors such as convergence to stable points, limit cycles and chaos \cite{yoshida1983analytic, strogatz2014nonlinear, may1976simple}.

In fluidic systems, droplet generation can be thought of as a discrete event, corresponding to the moment when a droplet breaks free from the bulk phase, making droplet generation well positioned to be studied as a discrete-time dynamical system. Yet, little work has explored this connection so far \cite{sessoms2010complex}, despite extensive studies of droplet generation either for technological purposes \cite{schwartz2004droplet,song2006reactions,teh2008droplet, schneider2013potential}, or fundamental physical understanding. With regards to the latter, there is a conceptual connection between discrete-time dynamical systems and the study of droplet pattern generation, which relates the size of the droplets to their spacings, often revealing asymmetries even at low Reynolds numbers under laminar flow \cite{thorsen2001dynamic,garstecki2006formation,link2004geometrically, garstecki2005oscillations}. Such patterns further enable self-organization phenomena where generated droplets are driven into ordered structures \cite{kita2008self,timonen2013switchable}.

Unlike microchannel configurations for droplet generation \cite{christopher2007microfluidic} that induce shearing between the two phases through T-junctions \cite{thorsen2001dynamic,garstecki2006formation, tice2003formation} or flow focusing \cite{anna2003formation,li2008simultaneous}, where the droplet formation timescales arises from balance of viscous forces and capillary pressure, in this work, we report a novel microfluidic system with an intrinsic driving frequency determined by the  time-varying magnetic energy landscape with a two-phase immiscible mixture of water-based ferrofluid (FF) and silicone oil. The magnetic energy landscape generates an oscillatory force that produces the droplet arrays whose patterns depend on the energy of breakup, the oscillation frequency and a bias flow-rate. The same concept of magnetic energy landscapes has previously been utilized to synchronously manipulate water-based FF droplets and, through droplet-to-droplet interactions, perform physical logic operations \cite{katsikis2015:sd}. In this letter, we use this platform to demonstrate control over periodic droplet patterns, characterized by different droplet-to-droplet spacing and droplet sizes, and develop a discrete-time dynamical systems model to explain the dynamics driving the formation of these patterns.

\section*{Experimental methods}
\label{sec:bulk_swimming}

\begin{figure}
\begin{center}
\includegraphics[width=1.0\textwidth]{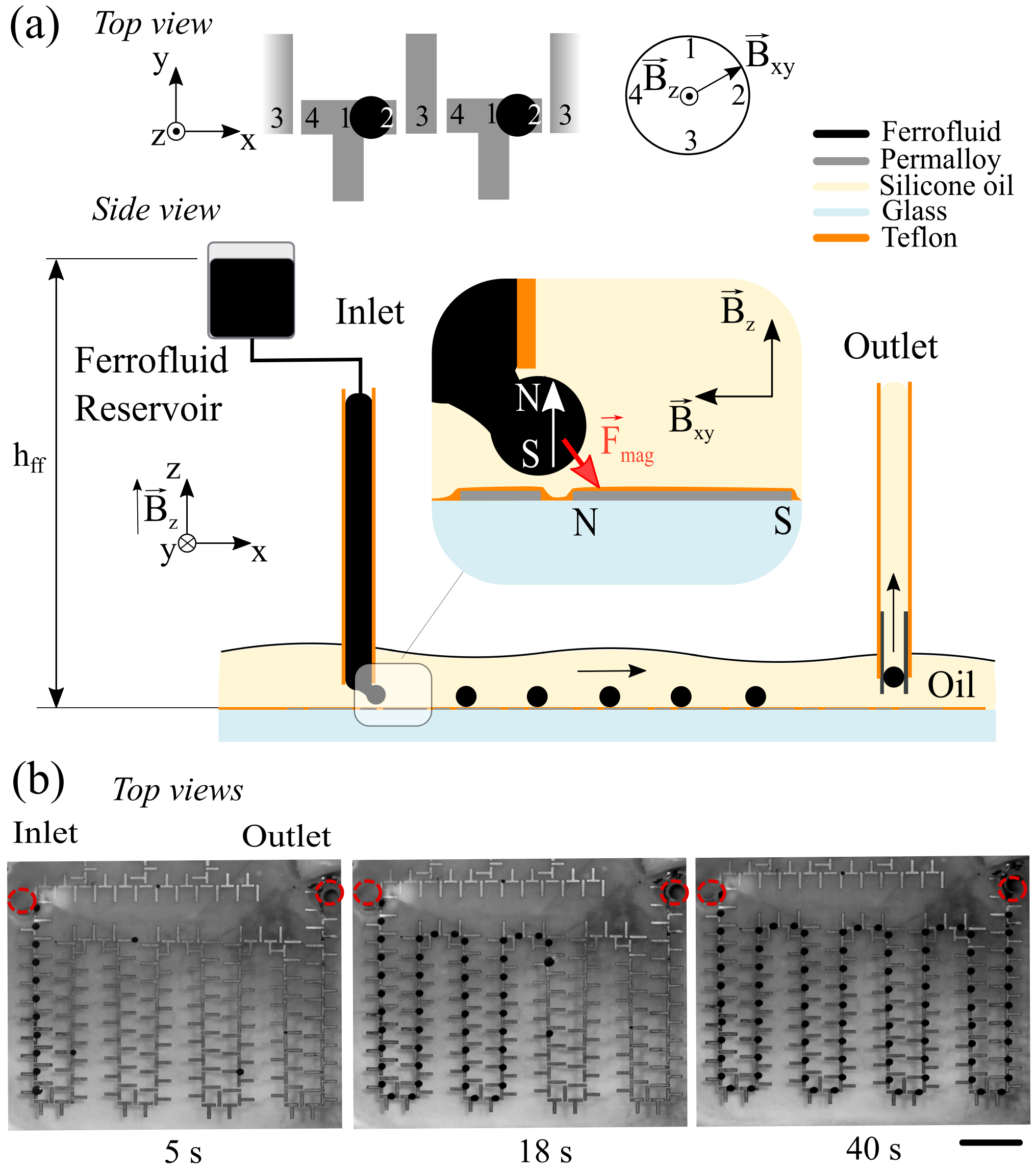} 
\caption{\label{fig:schematic}\textbf{Schematic of droplet generator and experiments} (a) Schematic of the droplet generator. \textit{Top view}: Periodic tracks of `T' and `I' permalloy bars (gray) with FF droplets (black) propagating under magnetic fields, ${B}_{z}$, ${B}_{xy}$. The numbers ``1-4" on the bars correspond to the locations that the droplets occupy as ${B}_{xy}$ obtains the angular orientations ``1-4" \cite{katsikis2015:sd}. \textit{Side view}: Droplet array generated from reservoir with height, $h_{ff}$, via a magnetic force ${F}_{mag}$ (red). The letters `N' and `S' denote polarizations. (b) Top-view sequential snapshots of generated droplets propagating on winding tracks of `T' and `I' bars. Red dashed circles indicate the inlet and outlet. ${B}_{z}=250~G$, ${B}_{xy}=40~G$ at frequency $f=2~Hz$. Scale bar $5~mm$.
}
\end{center}
\end{figure}

We supply the FF through an inlet tubing (diameter ${d}_{tube}=300~\mu m$) that is placed at a distance $d = 50-200~\mu m$ from a substrate covered with a $3-5~mm$ thick film of silicone oil (Fig. \ref{fig:schematic}a, side view; Supp. Information). The FF reservoir is held at a height $h_{ff}$ from the substrate, that creates a differential pressure $\Delta P={\rho}_{ff}~g~h_{ff}$, where ${\rho}_{ff}=1.28~g/{cm}^{3}$ is the density of the FF and $g=9.81~m/{s}^{2}$ is the acceleration of gravity. Due to this pressure, $\Delta P$, there is flow of bulk FF with a rate $Q$.

\begin{figure}
\begin{center}
\includegraphics[width=1.0\textwidth]{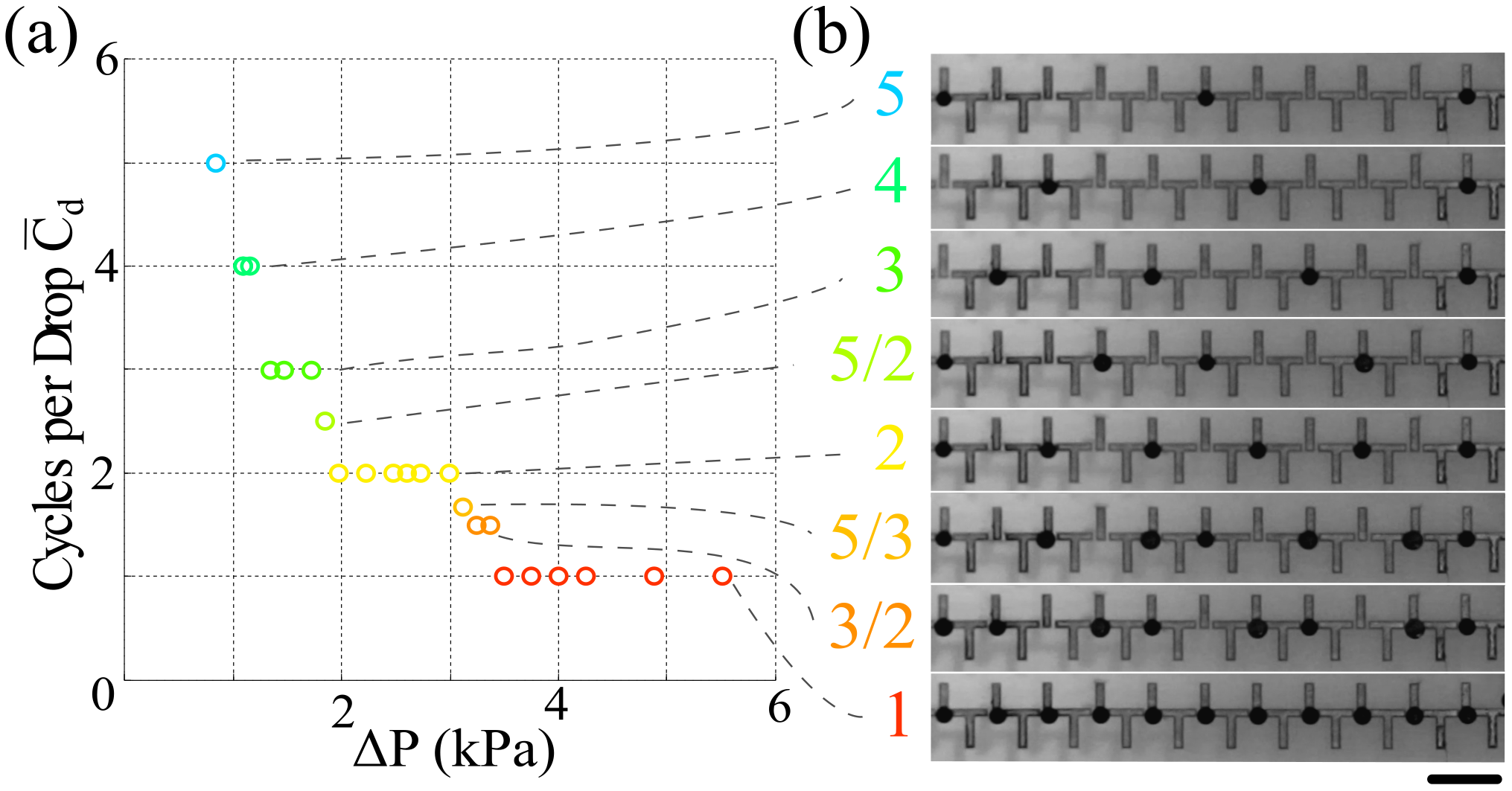}
\caption{\label{fig:dropletpatterns} (a) Plot of $\bar{C}_{d}$, the average number of cycles of ${B}_{xy}$ needed to generate a droplet, versus the hydrostatic pressure $\Delta P$. (b) Snapshots from the experiments of (a) with droplet-to-droplet spacings corresponding to different $\bar{C}_{d}$ values. ${B}_{z}=250~G$, ${B}_{xy}=40~G$, $f=2~Hz$. Scale bar $2~mm$.
}
\end{center}
\end{figure}

The droplets are generated through the interaction of the bulk FF with soft-magnetic (permalloy) tracks (characteristic length $\sim 1~mm$) on the substrate via exposure to two magnetic fields. The first magnetic field, $\mid{B}_{z}\mid=250~G$, is perpendicular to the substrate, has a fixed magnitude and polarizes the bulk FF in a uniform manner. The second magnetic field,  $\mid {B}_{xy} \mid=40~G$, is in the plane of the substrate, is rotating with a radial frequency $\omega$ and polarizes the tracks. As a result, these magnetic fields create a dynamic, spatiotemporal magnetic energy landscape, where the FF will be driven towards the minimization of its potential energy. To accomplish this, the lower end of the bulk is subject to a magnetic force $\overrightarrow{F}_{mag}$ that extracts sub-millimeter diameter droplets (Fig. \ref{fig:schematic}a, side view). For this study, we restrict ourselves to tracks that have shapes of `T' and `I' bars that ensure that they can be polarized effectively by the $\overrightarrow{B}_{xy}$ and suffice not only to generate droplets but also to propagate them along the tracks (Fig. \ref{fig:schematic}a, top view). For a fixed position of the inlet tube, we show both droplet generation and propagation along the tracks of the substrate (Fig. \ref{fig:schematic}b, top view; Supp. Video 1).  To avoid overcrowding the substrate with droplets, we use outlet lines connected to a negative pressure line that remove the droplets from the substrate (Fig. \ref{fig:schematic}a and S1). 

\begin{figure}
\begin{center}
\includegraphics[width=0.8\textwidth]{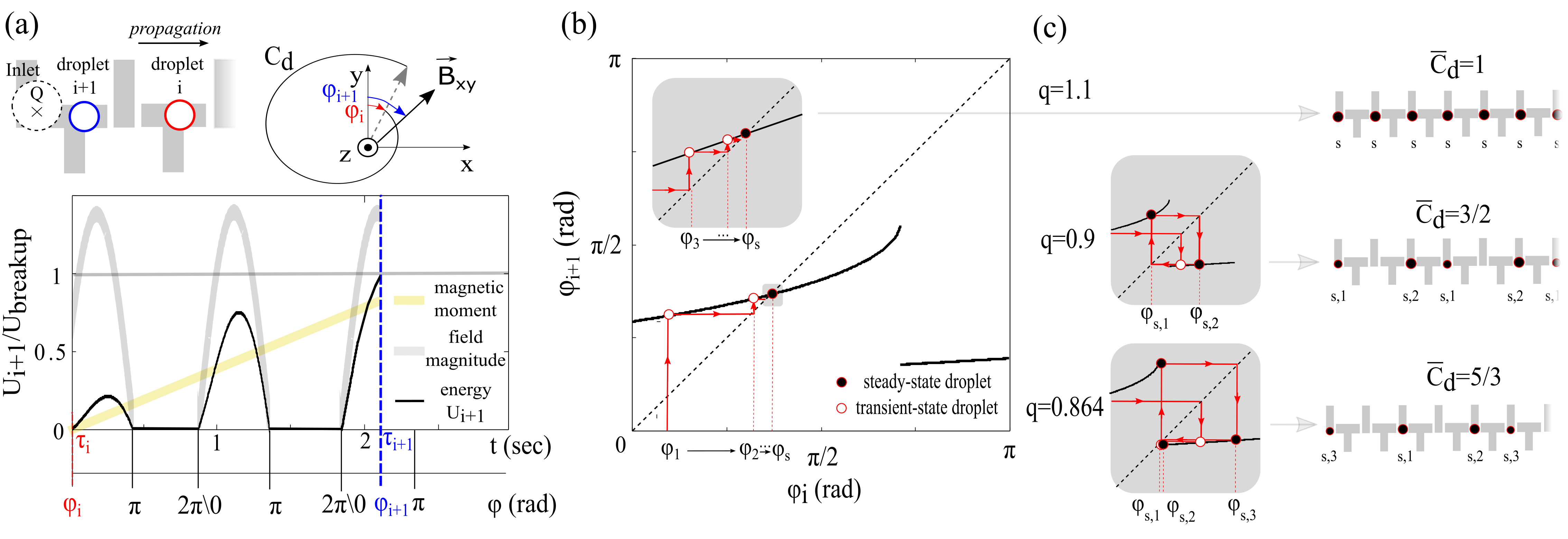} 
\caption{\label{fig:modelintro} (a) Schematic for recursive model. Assuming that the droplet `i' (red) is generated at time $t=\tau_{i}$ when ${B}_{xy}$ is at ${\phi}_{i}$, the next droplet `i+1' (blue) will be generated at $t=\tau_{i+1}$ and ${\phi}_{i+1}$, when the energy of droplet `i+1' becomes ${U}_{i+1}={U}_{breakup}$, after a number of cycles ${C}_{d}=1,2,..$ (black spiral), based on equation (\ref{model_eq_recursive}). ${U}_{i+1}$ (black) is the total product of the normalized magnetic moment $qt$ (yellow) and normalized magnetic field $f(t,~\varphi_i) = \textrm{max}(sin(2\pi t + \varphi_i),~0)$ (gray). (b) Plot of ${\phi}_{i+1}$ versus ${\phi}_{i}$ (solid black curve) based on the solution of equation (\ref{model_eq_recursive}) for q=1.1. The dashed line indicates the $y=x$ line and the red lines indicate the convergence of initial random ${\phi}_{1}$ of transient-state droplets (white circles), to a ${\phi}_{s}$ of steady-state droplets (black circle). The gray inset shows a zoomed-in graphical solution converging to ${\phi}_{s}$. (c) Graphical solutions of ${\phi}_{i+1}$ versus ${\phi}_{i}$ for $q=1.1,~0.9,~0.864$ corresponding to $\bar{C_d}=1, 3/2, 5/3 $ with respective illustrations of sizes and spacings of droplets.}
\end{center}
\end{figure}
%
\section*{Experimental observations}
\label{sec:experimental_observations}
%
For given magnetic fields and fixed positions of the inlet and outlet tubes, we apply pressures in the range $\Delta{P}=0.5-8~kPa$ and observe that the generated droplet arrays converge to a steady-state pattern within 2-3 cycles of $\overrightarrow{B}_{xy}$. The system is in a constant flow and pressure regime (Supplementary Information). In steady-state, there is a minimum of one full cycle of $\overrightarrow {B}_{xy}$ required to generate a single droplet ($C_d=1$). Here, we define $C_d$ as the number of cycles of rotating magnetic field per released drop. For decreasing $\Delta{P}$, more cycles are required for the generation of single droplet ($C_d\geq 1$), resulting in droplet arrays that are less tightly spaced (Fig. \ref{fig:dropletpatterns} and Supp. Video 2). Interestingly, we observe that the spacings between consecutive droplets can be non-constant, nonetheless still having a periodic pattern; for example there can be periodic alternation between one and two cycles per droplet (i.e. ${C}_{d}$ exhibits sequence `..1-2-1-2..'), resulting in an average of $\bar{C_d}=3/2$ (Fig. \ref{fig:dropletpatterns}b). In these cases, the volumes of the droplets can also be different.
%
\section*{Model}
\label{sec:model}
%

To explain the different droplet-to-droplet spacing and individual droplet volumes in our generated arrays (Fig. \ref{fig:dropletpatterns}b), we develop a theoretical model. We write a tractable expression for the magnetostatic energy of the droplet, which theoretically is defined as $U = \int{ -\overrightarrow{M} \cdot \overrightarrow{B}_{bar} dV}$, where $\overrightarrow{M}$ is the magnetization of the droplet, $\overrightarrow{B}_{bar}$ is the magnetic field generated by the bars and $V$ is the volume of the droplet. 
To simplify the complicated expression for $U$ (Supplementary Information), we base our model on the following five assumptions: First, we consider the droplet as a point mass and write $U =- \overrightarrow{M}\cdot \overrightarrow{B}_{bar} V$. Second, we assume that $\overrightarrow{M}=M\hat{z}$ (Fig. \ref{fig:schematic}a) with $V$ increasing linearly over time $t$ for a given flow rate $Q$, allowing us to write the magnitude of the magnetic moment $\overrightarrow{\mu}=\overrightarrow{M}V$ as $\mu(t) = M_d Q t$. Third, we assume that $\overrightarrow{B}_{bar}={B}_{bar}\hat{z}$ with ${B}_{bar}$ varying as a sine wave over time, consistent with the oscillatory nature of $\overrightarrow{B}_{xy}$, and thus write ${B}_{bar}(t)= B_0sin (\omega t + \varphi_i)$ where $B_0$ is positive and is the maximum amplitude of $\overrightarrow{B}_{bar}$, $\omega$ is the angular frequency, and $\varphi_i$ is the phase of $\overrightarrow{B}_{xy}$. Fourth, we assume that a droplet breaks up from the bulk when its energy $U$ is minimized to a threshold $U_{breakup}$ which is constant and does not depend on droplet volume. We base this assumption on the fact that, for droplets that are roughly the diameter of the inlet tube or larger, $U_{breakup}$ is set by the cross-sectional area of the inlet tube and the surface energies of the fluids. In our experiments, the radius of the smallest droplet was measured to be $r_{min} = 240 \mu m$, suggesting a constant $U_{breakup}$. Additionally, for the rest of this work, we will refer to the absolute value of the energy $U$. Fifth, we assume that droplet breakup can occur only in the attractive phase of the oscillation when $sin(wt + \varphi)>0$ and ${B}_{bar}(t)>0$. In the repulsive phase, the droplet is pushed away from the magnetized bar, which then reduces the applied magnetic force on the droplet, preventing breakup from occurring. 

Combining all five of these assumptions, we write down the equation for the magnetostatic energy of the model as:

\begin{figure}
\begin{center}
\includegraphics[width=0.8\textwidth]{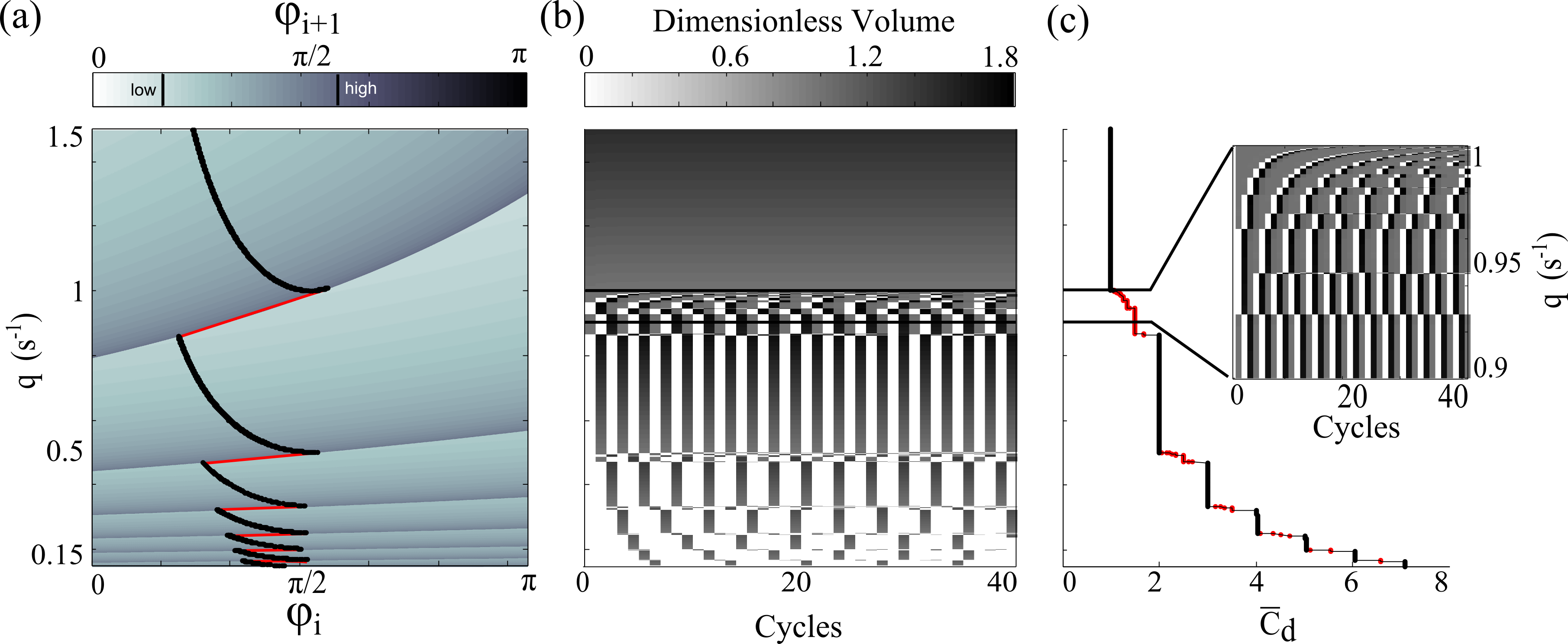}
\caption{\label{fig:model_param_space} Simulation parameter sweep of equation (\ref{model_eq_recursive}). (a) Phase map where each row corresponds to a mapping from $\varphi_i$ (x-axis) to $\varphi_{i+1}$ (colorbar), for a given flow rate $q$. The black lines correspond to steady-state points where phase maps intersect the unity lines with positive slope at exactly one point. In these domains, droplet generation will be monodisperse. Red lines denote regions of multiple steady-state points for $\varphi_i \rightarrow \varphi_{i+1}$ mapping. Low and high bounds in colorbar correspond to mapping limits given any initial $\varphi_i$. (b) Plot of droplet volumes for discrete cycles of ${B}_{xy}$ as a function of flow rate ($q$). White cells indicate cycles where no droplet was released. Cell shade indicates dimensionless droplet volume at a given cycle (colorbar). (c) Plot of $\bar{C_d}$ as a function of $q$. Red dots correspond to regions of multiple stead-state points as in (a).}
\end{center}
\end{figure}

\begin{equation}
U(t) = \begin{cases} B_0 M_d Q t sin(wt + \varphi) & ,~sin(wt + \varphi) \geq 0 \\ 0 & ,~sin(wt + \varphi)<0 \end{cases}
\label{model_continuous_eq}
\end{equation} 

Once a droplet is released, only the phase of $\overrightarrow{B}_{xy}$ at the previous breakup is needed to determine the time to next breakup. This allows us to write equation (\ref{model_continuous_eq}) as a recursive formula; assuming that a droplet `i' is generated at time $t=\tau_{i}$ when $\overrightarrow{B}_{xy}$ is at angle ${\varphi}_{i}$, then the next droplet `i+1' will be generated at time $t=\tau_{i+1}$ and ${\phi}_{i+1}$, which occurs when the droplet magnetic energy is equal to $U({\tau}_{i+1})={U}_{i+1}={U}_{breakup}$ (Fig. \ref{fig:modelintro}b). Without loss of generality, we reduce equation (\ref{model_continuous_eq}) by setting $B_0M_dQ = q$ (${s}^{-1}$), $\omega=2\pi$ ($rad/sec$) and ${U}_{breakup}=1$, and write the recursive expression as:

\begin{equation}
	q \tau_{i+1} f(\tau_{i+1},~\varphi_i) = 1
    \label{model_eq_recursive}
\end{equation}

\begin{figure}
\begin{center}
\includegraphics[width=0.455\textwidth]{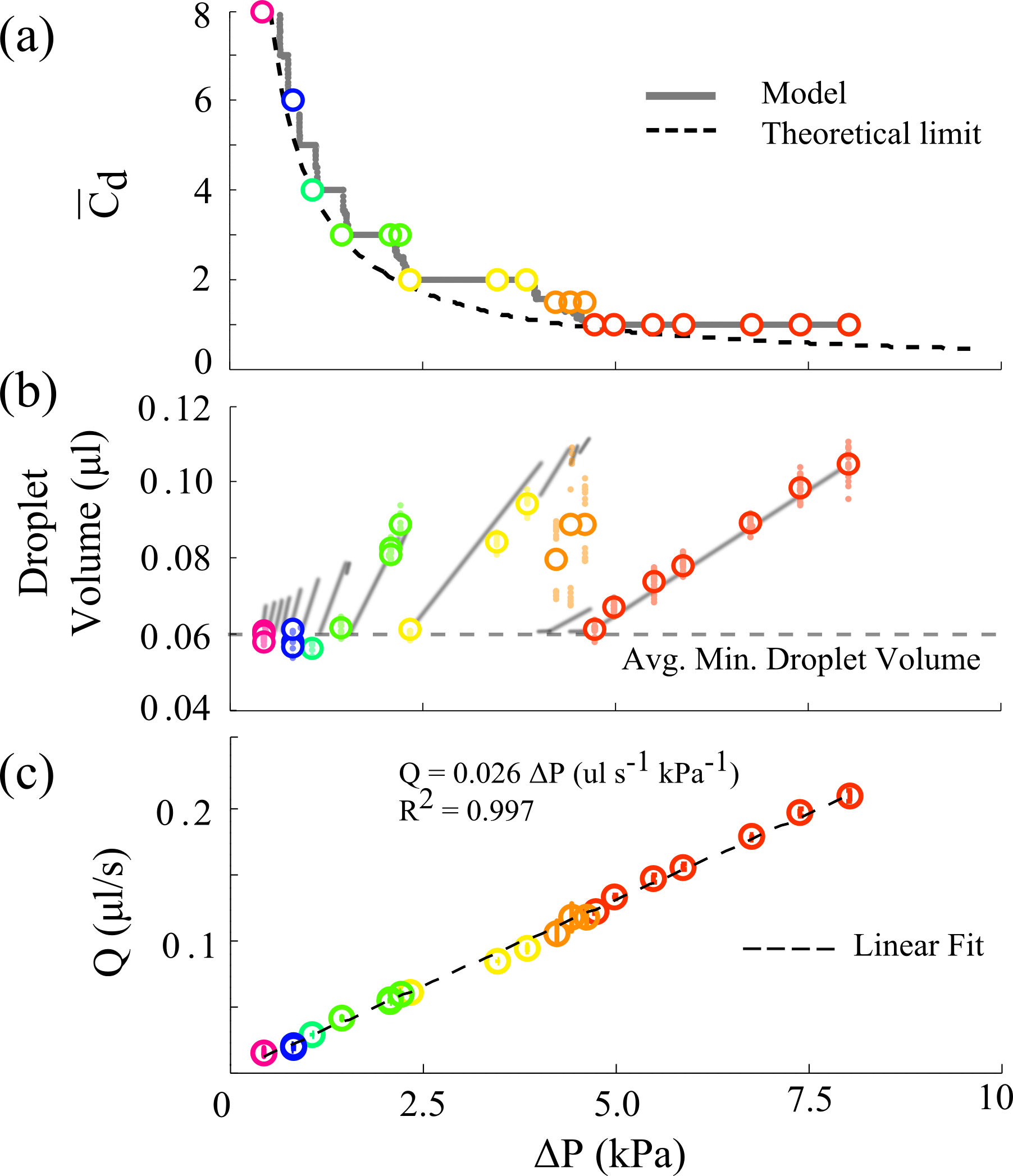}
\caption{\label{fig:experiments} Experimental droplet generation. (a)Plot of $\bar{C_d}$, over pressure $\Delta P$. Colors serve as a legend for panels (b) and (c). Dashed black line is the theoretical minimum of $\bar{C_d}$ given the average minimum droplet volume ($V_{min} = 0.59 \mu l$). Gray line is fit of $\bar{C_d}$ from the solution of equation (\ref{model_eq_recursive}) given a single-parameter fit using $V_{min}$. (b) Droplet volumes as a function of pressure. Large colored circles are average values for generated droplet volumes at a given pressure. Smaller colored dots correspond to individual droplet volumes. Dashed horizontal line is the average minimum droplet volume. Gray lines are values of droplet volumes from model solutions given $V_{min}$. (c) Plot of Flow rate over $\Delta P$ with linear fit.}
\end{center}
\end{figure}

Where $f$ is the waveform of the magnetic field relevant for breakup, and is given by $f(t,~\varphi_i) = \textrm{max}(sin(2\pi t + \varphi_i),~0)$ (Fig. \ref{fig:modelintro}a, gray field magnitude curve). Next, we solve equation (\ref{model_eq_recursive}) to reproduce the phenotype of the droplet arrays generated experimentally (Fig. \ref{fig:dropletpatterns}b). For given $q$ and angles $\varphi_{i}$ in the range $[0,\pi]$, we find the corresponding values of $\tau_{i+1}$. We restrict our parameter range for $\varphi_i$ to an upper bound of $\pi$ since no breakup can occur from $\pi$ to $2\pi$. Then, we calculate both the angle $\varphi_{i+1}$ based on equation $\varphi_{i+1} = \textrm{mod}_{2\pi} (2\pi \tau_{i+1} + \varphi_i)$ and the number of cycles $C_{d}$ required to generate a droplet `i+1' based on $C_{d} = \textrm{quotient}_{2\pi} (2\pi \tau_{i+1} + \varphi_i)$, therefore generating phase maps for specific $q$ values that relate $\varphi_i$ to $\varphi_{i+1}$ (Fig. \ref{fig:modelintro}b,c).  For $q=1.1$, $\varphi_{i}$ converges to a single steady-state angle $\varphi_{s}$  (Fig. \ref{fig:modelintro}b) resulting in monodisperse droplets with $\bar{C}_{d}=1$ (Fig. \ref{fig:modelintro}c) independent of the initial $\varphi_{0}$. In other regimes, for example at $q=0.9$ and $q=0.864$, $\varphi_{i}$ periodically alternates respectively between two and three steady-state angles (Fig. \ref{fig:modelintro}c) resulting in multidisperse droplets with $\bar{C}_{d}=3/2$ and $\bar{C}_{d}=5/3$ in qualitative agreement with experiments (Fig. \ref{fig:dropletpatterns}b).

To study the stability and pattern-space of the model, we conduct a parameter sweep of $q$ in the range: $[0.15, 1.5]$ (Fig. \ref{fig:model_param_space}). The phase-stability map reveals domains of single steady-state points, where $\varphi_i = \varphi_{i+1}$ with integer $\bar{C_d}$ values (Fig. \ref{fig:model_param_space}a; black lines), that are interrupted by domains of multiple steady-state points and non-integer $\bar{C_d}$ values (Fig. \ref{fig:model_param_space}a; red lines). These multidisperse transitional domains occur at discontinuous boundaries in the phase map (qualitatively as in Fig. \ref{fig:modelintro}c, $\bar{C_d} = 3/2, ~5/3$). In addition, for the explored parameter range, we find that given any initial $\varphi_i$ value, the subsequent $\varphi_{i+1}$ is always narrowed to a band of $[0.509, ~1.771] ~ rad$ (Fig. \ref{fig:model_param_space}a, colorbar; Supp. Info).

Furthermore, to illustrate the richness in potential droplet spacing and volume patterns, we calculate droplet volume over discrete cycle intervals at different $q$ values (Fig. \ref{fig:model_param_space}b). The pattern-space includes monodisperse and multidisperse droplet sequences at $\bar{C_d}$ (Fig. \ref{fig:model_param_space}c) values observed experimentally (Fig. \ref{fig:dropletpatterns}c).

\section*{Comparison of experiment and model}
To understand the relationship between the droplet volume and pressure, we study one configuration at an in-plane frequency of $2~Hz$, describe the measured physical quantities in detail and test our analytic model by comparing to the experimental results (Fig. \ref{fig:experiments}). 

Decreasing pressure down from $8 ~kPa$, we find monotonically increasing $C_d$ values (Fig. \ref{fig:experiments}a). For a given $\bar{C_d}$ value, average droplet volume decreases with decreasing pressure. As $\bar{C_d}$ transitions from 1 to 2, 2 to 3/2 and 3/2 to 3, droplet volumes jump abruptly to higher values before decreasing again (Fig. \ref{fig:experiments}b). We find that the average minimal droplet volume for all integer $\bar{C_d}$ is $V_{min} = 0.059 \mu l$ (Fig. \ref{fig:experiments}b, dashed line). Plotting the flow rate, $Q = V_{drop} f/\bar{C}_d$, as function of $\Delta P$ gives a linear relationship with a slope of $26.2 * 10^{-3} \, \mu l \, s^{-1} \, kPa^{-1}$ (Fig. \ref{fig:experiments}c, $R^2 = 0.997$). The linearity of this relationship confirms that the magnetic, capillary and hydrodynamic forces at the exit of the tube are much smaller than the force driving the FF flow.

Given the experimentally determined $V_{min}$, we can reevaluate equation (\ref{model_eq_recursive}) and compare theory to experiment, by parameterizing $q = Q/V_{min}$ and setting $w = 2*2\pi$. We use the recursive equation (\ref{model_eq_recursive}) to numerically solve exact values of $\bar{C_d}$ (Fig. \ref{fig:experiments}a, gray line) and the droplet volumes (Fig. \ref{fig:experiments}b) for $V_{min} = 0.059 ~ \mu l$ over a range of $q$. With $V_{min}$ as the single-parameter fit, we find good qualitative agreement between experiment and theory, particularly in the transitions between different $\bar{C_d}$. For $\bar{C_d} = 3/2$, we find a difference in expected droplet volumes suggesting that there may need to be important corrections made to the $B_{xy}$ waveform.
%
\section*{Conclusions}
In summary, we have demonstrated an experimental platform in which a periodic force generates droplet arrays with complex patterns of droplet spacings and sizes. We have developed a discrete-time dynamical systems model to explain the observed patterns, and found good agreement with experimental measurements. More broadly, this work may suggest a new formalism to study droplet generation under time-dependent force using iterative phase maps and other discrete-time dynamical systems approaches.
%
 \section*{Acknowledgements}
The authors thank all members of the Prakash Lab for useful discussions. AR acknowledges support from the NSF GRFP. GK acknowledges support from the Onassis Foundation and A. G. Leventis Foundation. MP acknowledges support from Pew Foundation and Keck Foundation.

\pagebreak
\pagebreak
\begin{center}
\textbf{\large Supplemental Materials: Title for main text}
\end{center}

\section{Experimental methods}

\textit{Fabrication of fluidic chips $-$} The `T' and `I' bars are fabricated by etching permalloy foils that are epoxy-bonded on glass substrates, using an protocol identical to reference \cite{Katsikis2015}. The `T' and `I' bars (Fig.S1) have millimeter-size dimensions (Table \ref{Table_dimensions}). The permalloy bars are coated with teflon and the fluidic chips do not have a top cover. 

\begin{table}[h!]
\centering
\begin{tabular}{ |c|c|} 
\hline
\multicolumn{2}{|c|}{Dimensions of `T' and `I' bars}
\\
\cline{1-2}
\includegraphics[scale=0.4]{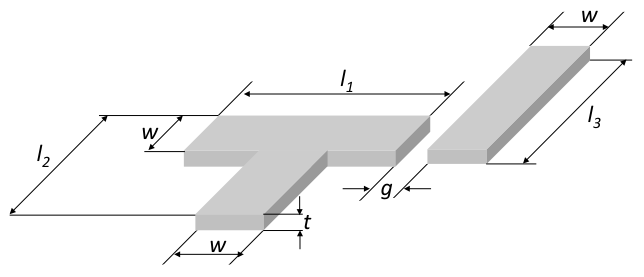}
& ($\mu m$)
\\ 
\hline
${l}_{1}$ & 1012.5 \\ 
\hline
${l}_{2}$ & 1125 \\ 
\hline
${l}_{3}$ & 1125 \\ 
\hline
${w}$ & 213.5 \\ 
\hline
${t}$ & 25 \\ 
\hline
${g}$ & 70 \\ 
\hline
\end{tabular}
\caption{Table with nominal dimensions of `T' and `I' bars}
\label{Table_dimensions}
\end{table}

\textit{Magnetic fields $-$} The magnetic fields are generated using the system of electromagnetic coils described in reference \cite{Katsikis2015}. The ratio between the magnitudes of the magnetic fields is $\mid {B}_{z} \mid/ \mid {B}_{xy} \mid \geq 5$, thus ensuring that the induced magnetization of the generated droplets is along the z-axis (Fig. \ref{fig:schematic}A). However, the induced magnetization of the metallic bars is
always in the x-y plane as they are too thin (for example ${t}/{l}_{1}=1/40$) to support magnetization in the z-axis.

\textit{Two-phase mixture of fluids $-$} The mixture consists of two phases. The first phase is silicone oil (Sigma Aldrich, CAS: 63148-62-9, kinematic viscosity $5~cSt$, density $0.913~kg/{m}^3$) which is pipeted on the surface of the fluidic chip forming a film of thickness $h_{oil} = 3-5mm$ beneath the open air-oil interface.  The second phase is water-based ferrofluid (Ferrotec EMG 700, kinematic viscosity $5~cSt$, density $1.28~kg/{m}^3$) which is dispensed on the film using an inlet tubing (Fig. \ref{fig:schematic}A).

\textit{Inlet tubing $-$} The inlet tubing is made of teflon (PTFE) with internal diameter  $300~ \mu m$ and length $1~m$. The first tip of the tubing is suspended at a height $d = 50-200 ~ \mu m$ above the permalloy bars. This height $d$ is always smaller than the thickness of the silicone oil film on the substrate, that is $d<{h}_{oil}$, thus making this tip completely immersed in the film. The second tip of the tubing is connected to a ferrofluid reservoir whose top surface is at a height $h_{ff}=10-80~mm$ above the permalloy bars. This height $h_{ff}$ creates a pressure difference $\Delta P$ generating flow that fills the tubing with ferrofluid and - via the first tip -  dispenses it into the substrate. The pressure difference $\Delta P$ is adjusted by adjusting the height  $h_{ff}$ of the ferrofluid reservoir. Furthermore, the inlet tubing is threaded through a glass capillary with internal diameter $500 \mu m$, which is mounted on a three-axis translational stage for adjusting the position of the end of the inlet tubing relative to the permalloy bars. The height of the oil ($h_{oil}$) contributes an insignificant reduction in pressure and is not considered here.

\begin{figure}[h!]
  \centering
      \includegraphics[scale=0.925]{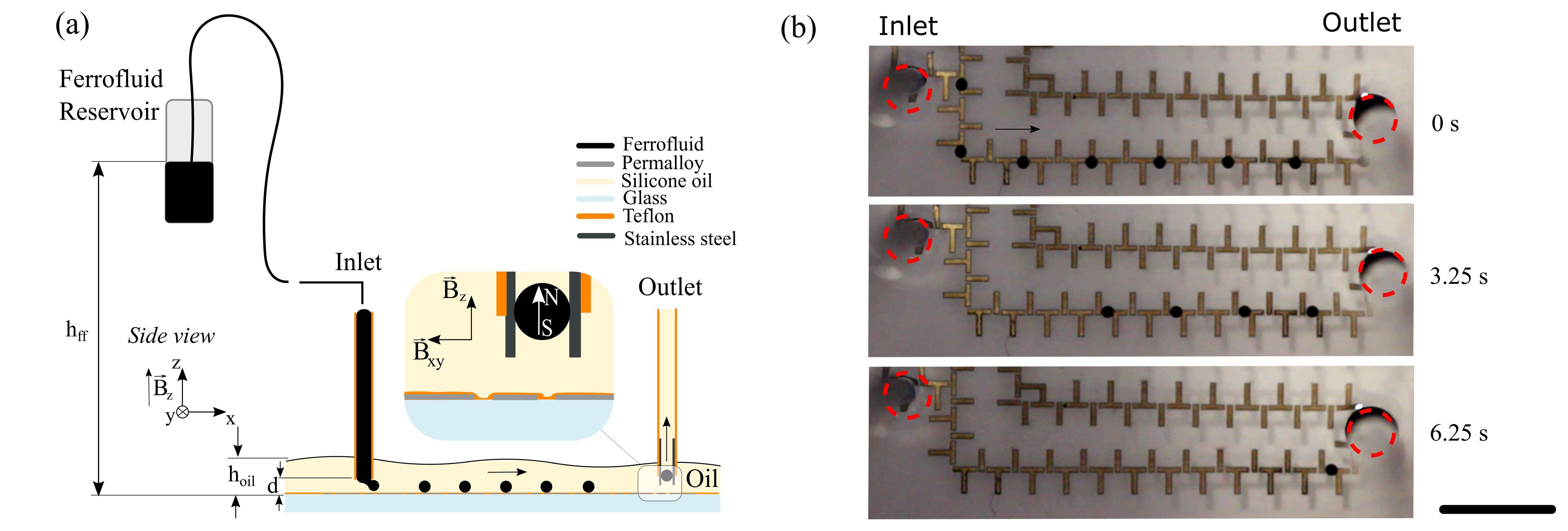}
  \caption{(a) Schematic of droplet generator. \textit{Side view}: The inlet tube contains a FF column (black) with controllable hydrostatic pressure set by the height, $h_{ff}$. The droplets propagate on the tracks covered with silicone oil with height $h_{oil}$ and exit the substrate through an outlet tube (shown in inset) connected to a negative pressure line. (b) Top-view sequential snapshots of a experiment where generated droplets propagate on winding tracks of `T' and `I' bars and are removed from the substrate through the outlet. Red dashed circles indicate the inlet and outlet. ${B}_{n}=250~G$, ${B}_{i}=40~G$ at frequency $f=2~Hz$. Scale bar $5 mm$.}
\label{fig:trash_line}
\end{figure}

\textit{Outlet tubing $-$} The outlet tubing is made of teflon, similar to the inlet tubing (Fig. \ref{fig:trash_line}a). At its lower end that is in proximity to the substrate, it also contains a blunt-tip pin made of stainless steel (23 gauge). The magnetic field $\overrightarrow{B}_{z}$ along the z-axis magnetizes the pin. The magnetized pin attracts the ferrofluid droplets and by also using an additional negative pressure difference across the outlet tubing, the droplets that reach the outlet tubing are removed from the substrate (Fig. \ref{fig:schematic} b and Fig. \ref{fig:trash_line}b).

\textit{Imaging $-$} Droplet volume measurements are performed by imaging the chip with a dSLR (Canon T3i, Canon EF 100mm f/2.8L IS USM Macro Lens).

\begin{figure}[h!]
  \centering
      \includegraphics[width=0.33\textwidth]{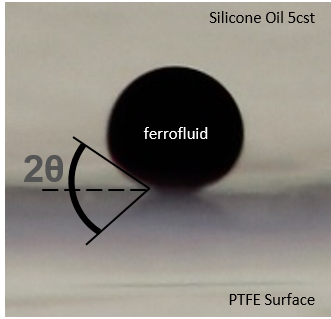}
  \caption{A sessile ferrofluid droplet resting on a glass-PTFE spun coat surface in 5cst silicone oil. Image taken from the side. The droplet radius is approximately $300 \mu m$.}
  \label{fig:sessile}
\end{figure}

\textit{PTFE-Oil-Ferrofluid Surface Energy}. In order to estimate the volume of sessile droplets, by only imaging from the top, we measured the contact angle between ferrofluid, PTFE in silicone oil. We measured 11 droplets from the side, sessile on a PTFE surface, for an average surface angle of $\theta = 24.86 \pm 2.72$ (Fig. \ref{fig:sessile}).

\section{Data Analysis} \label{app:foobar}

\textit{Droplet Volume Measurement}. For each measurement, droplets are first generated and then all B-fields are turned off, so that the droplets are in a sessile state on the chip surface. Droplet radii are measured using contrast-based object detection in Matlab ( {\color{orange}{YORGOS - add any comments here}}). For maximal droplet volumes of $V_{droplet} \approx 0.12 ul$, the Bond number is $\approx 0.25$ ($\Delta \rho = \rho_{ff} - \rho_{oil} = 0.2 g/ml$; $\gamma = 3mN/m$ \cite{Flament1996}), therefore justifying the spherical cap assumption in calculating the volumes of the droplets, where the $V_{cap}(r,\theta) =  (\pi r^3 /  6) (1-cos{\theta})(3 sin{\theta}^2 + (1-cos{\theta})^2)$. In our system, the FF-teflon-oil surface contact angle is measured to be $\theta \approx 25 \degree$.

\section{Droplet Generation Videos}
\textbf{\textit{Video 1}}. Monodisperse droplet generation at $B_{xy}$ frequency of $2Hz$.

\textbf{\textit{Video 2}}. Pressure sweep of  droplet generation at $B_{xy}$ frequency of $2Hz$, displaying various $\bar{C_d}$ regimes.

\subsection{Model and Fits}

\textit{Computational Solution}. MATLAB R2014a was used to numerically solve the recursive equation (\ref{model_eq_recursive}). The recursive process is as follows: after the i-th droplet is generated, time is reset to $t = 0$ and $\varphi_i$ is propagated to the subsequent iteration. We next solve for the time, $\tau_{i+1}$, that it takes for the energy to reach $U_{breakup}$. To plot phase maps, we solve the recursive equation for a range of $\varphi$ from $0$ to  $\pi$ in increments of at least $0.001$.

\textit{Fitting}. Linear fitting was done using the first-order Polyfit function in Matlab. $R^2$ value was then calculated as an estimator of linearity.

\end{document}